\documentclass[twocolumn,english]{revtex4-2}
\usepackage[T1]{fontenc}
\usepackage[latin9]{luainputenc}
\usepackage{amsmath}
\usepackage{babel}
\begin{document}
\title{Four center integrals for Coulomb interactions in small molecules}
\author{Garry Goldstein}
\address{garrygoldsteinwinnipeg@gmail.com}
\begin{abstract}
In this work we make some progress on studying four center integrals
for the Coulomb energy for both Hartree Fock (HF) and Density Functional
Theory (DFT) calculations for small molecules. We consider basis wave
functions of the form of an arbitrary radial wave function multiplied
by a spherical harmonic and study four center Coulomb integrals for
them. We reformulated these Coulomb four center integrals in terms
of some derivatives of integrals of nearly factorable functions which
then depend on the Bessel transform of the radial wave functions considered.
\end{abstract}
\maketitle

\section{\protect\label{sec:General-considerations}Introduction}

Small molecules water, carbon dioxide, ozone, ammonia, methane, ethane,
ethene, ethyne to name a few have tremendous practical industrial
applications so it is paramount that we study their electronic structures
\citep{Blinder_2019,Helgaker_2000,Marx_2009}.To make progress with
the many body Schrodinger equation for a small molecule a variety
of different approximations need to be made. The most common ones
for abinitio studies of electronic structures of small molecules are
Hartree Fock (HF) or Density Functional Theory (DFT) \citep{Blinder_2019,Helgaker_2000,Martin_2020,Marx_2009}
approximations (where DFT is then further approximated by the Local
Density Approximation (LDA) or Generalized Gradient Approximation
(GGA) \citep{Martin_2020,Marx_2009}). As such it is of paramount
importance to make progress in the study of small molecules both through
Density Functional Theory (DFT) and Hartree Fock (HF) methods. One
of the stumbling blocks towards implementing DFT or HF calculations
on modern computers is the choice of the basis set. Indeed a significant
practical improvement in the efficiency of solution of the HF equations
was given by Roothaan \citep{Roothaan_1951} who mapped the HF integro-differential
equations into a system of linear equations and unknowns using basis
sets.Similarly the Khon Sham (KS) equations are a linear system of
equations when considered within a fixed basis set \citep{Martin_2020,Marx_2009}.
Recently substantial progress has been made \citep{Goldstein_2024(2)}
where it was argued that it is sufficient to consider basis sets of
the form given by:
\begin{equation}
\varphi_{\alpha}\left(\mathbf{r}\right)=Y_{l_{\alpha}m_{\alpha}}\left(\widehat{\mathbf{r}-\mathbf{R}_{\alpha}}\right)R_{l_{\alpha}}\left(\mathbf{r}-\mathbf{R}_{\alpha}\right)\label{eq:Wave_function}
\end{equation}
Here $Y_{lm}$'s are spherical harmonics, $\mathbf{R}_{\alpha}$ is
the position of a nucleus of an atom in the molecule and $R_{l}$
are arbitrary radial wave functions. In Ref \citep{Goldstein_2024(2)}
arguments were made as to the form of the optimal set of $R_{l}$'s.
Inspired by recent progress in basis sets for small molecules \citep{Goldstein_2024(2)}
we study four center Coulomb integrals for specific types of wave
functions given by Eq. (\ref{eq:Wave_function}) and in particular
or those considered in Ref. \citep{Goldstein_2024(2)}. Specifically
we are interested in integrals of the form: 
\begin{equation}
I_{abcd}=\int d^{3}\mathbf{r}_{1}\int d^{3}\mathbf{r}_{2}\frac{1}{\left|\mathbf{r}_{1}-\mathbf{r}_{2}\right|}\varphi_{a}^{*}\left(\mathbf{r}_{1}\right)\varphi_{b}\left(\mathbf{r}_{1}\right)\varphi_{c}^{*}\left(\mathbf{r}_{2}\right)\varphi_{d}\left(\mathbf{r}_{2}\right)\label{eq:configuration}
\end{equation}
which show up in calculations of the Coulomb energy both in DFT and
HF methods. Previously evaluation of four center integrals was limited
to $R_{l}$'s being Gaussians \citep{Boys_1950,Helgaker_2000} which
limited accuracy for small basis set sizes \citep{Helgaker_2000}.
Evaluating these integrals for a large number of wave functions, with
more complex structure then Gaussian, needed for an accurate basis
is often the bottleneck in abinitio molecular electronic structures
calculations \citep{Blinder_2019,Helgaker_2000}. In this work we
write these integrals in terms of derivatives of an integral in a
nearly factorable form. The integrals depended on the Bessel transform
of the radial wave function given by: 
\begin{equation}
\mathcal{R}_{l}\left(U\right)=\int_{0}^{\infty}\frac{\pi}{2U^{2}}drJ_{l}\left(Ur\right)R_{l}\left(r\right)r^{2}\label{eq:Bessel_transform-1}
\end{equation}
as well as auxiliary integrals. This should lead to further abinitio
studies of small molecules with basis sets more accurate then Gaussian.

\section{\protect\label{subsec:Setup}Main calculation}

We now calculate the integral considered in Eq. (\ref{eq:configuration}).

\subsection{\protect\label{subsec:Fourier-transform}Fourier transform}
\begin{widetext}
We begin by Fourier transforming the integral in Eq. (\ref{eq:configuration}).
Now we have that 
\begin{equation}
\frac{1}{\left|\mathbf{r}_{1}-\mathbf{r}_{2}\right|}=\int\frac{d^{3}\mathbf{K}}{\left(2\pi\right)^{3}}\frac{4\pi}{\mathbf{K}^{2}}\exp\left(i\mathbf{K}\cdot\left(\mathbf{r}_{1}-\mathbf{r}_{2}\right)\right)\label{eq:Coulomb}
\end{equation}
Furthermore we recall that \citep{Louks_1967}
\begin{equation}
\exp\left(i\mathbf{K}_{\alpha}\cdot\mathbf{r}\right)=\exp\left(i\mathbf{K}_{\alpha}\cdot\mathbf{R}_{\alpha}\right)\sum_{lm}i^{l}J_{l}\left(\left|\mathbf{K}_{\alpha}\right|\left|\mathbf{r}-\mathbf{R}_{\alpha}\right|\right)Y_{lm}^{*}\left(\hat{\mathbf{K}}_{\alpha}\right)Y_{lm}\left(\widehat{\mathbf{r}-\mathbf{R}_{\alpha}}\right)\label{eq:Exponential}
\end{equation}
so that 
\begin{equation}
\varphi_{\alpha}\left(\mathbf{r}\right)=\left(-i\right)^{l_{\alpha}}\int\frac{d^{3}\mathbf{K}_{\alpha}}{\left(2\pi\right)^{3}}\exp\left(-i\mathbf{K}_{\alpha}\cdot\mathbf{R}_{\alpha}\right)Y_{l_{\alpha}m_{\alpha}}\left(\hat{\mathbf{K}}_{\alpha}\right)\int_{0}^{\infty}\frac{\pi\left|\mathbf{K}_{\alpha}\right|^{2}}{2}dr_{\alpha}J_{l_{\alpha}}\left(\left|\mathbf{K}_{\alpha}\right|r_{\alpha}\right)R_{l_{\alpha}}\left(r_{\alpha}\right)r_{\alpha}^{2}\exp\left(i\mathbf{K}_{\alpha}\cdot\mathbf{r}\right)\label{eq:Fourier}
\end{equation}
Here we have used orthogonality of Bessel functions (the Bessel transform
\citep{Shankar_1994}).

\subsection{\protect\label{subsec:Fourier-transformed-integral}Fourier transformed
integral}

As such we have that: 
\begin{align}
I_{abcd} & =\int d^{3}\mathbf{r}_{1}\int d^{3}\mathbf{r}_{2}\int\frac{d^{3}\mathbf{K}}{\left(2\pi\right)^{3}}\frac{d^{3}\mathbf{K}_{a}}{\left(2\pi\right)^{3}}\frac{d^{3}\mathbf{K}_{b}}{\left(2\pi\right)^{3}}\frac{d^{3}\mathbf{K}_{c}}{\left(2\pi\right)^{3}}\frac{d^{3}\mathbf{K}_{d}}{\left(2\pi\right)^{3}}\times\nonumber \\
 & \times\frac{4\pi}{\mathbf{K}^{2}}\exp\left(i\left(\mathbf{K}_{b}-\mathbf{K}_{a}+\mathbf{K}\right)\cdot\mathbf{r}_{1}\right)\exp\left(i\left(\mathbf{K}_{d}-\mathbf{K}_{c}-\mathbf{K}\right)\cdot\mathbf{r}_{2}\right)\times\nonumber \\
 & \times\left(-i\right)^{l_{b}+l_{d}}i{}^{l_{a}+l_{c}}\exp\left(i\left[\mathbf{K}_{a}\cdot\mathbf{R}_{a}+\mathbf{K}_{c}\cdot\mathbf{R}_{c}-\mathbf{K}_{b}\cdot\mathbf{R}_{b}-\mathbf{K}_{d}\cdot\mathbf{R}_{d}\right]\right)\times\nonumber \\
 & \times Y_{l_{a}m_{a}}^{*}\left(\hat{\mathbf{K}}_{a}\right)Y_{l_{c}m_{c}}^{*}\left(\hat{\mathbf{K}}_{c}\right)Y_{l_{b}m_{b}}\left(\hat{\mathbf{K}}_{b}\right)Y_{l_{d}m_{d}}\left(\hat{\mathbf{K}}_{d}\right)\times\nonumber \\
 & \times\int_{0}^{\infty}dr_{a}\frac{\pi\left|\mathbf{K}_{a}\right|^{2}}{2}J_{l_{a}}\left(\left|\mathbf{K}_{a}\right|r_{a}\right)R_{l_{a}}\left(r_{a}\right)r_{a}^{2}\times\int_{0}^{\infty}dr_{b}\frac{\pi\left|\mathbf{K}_{b}\right|^{2}}{2}J_{l_{b}}\left(\left|\mathbf{K}_{b}\right|r_{b}\right)R_{l_{b}}\left(r_{b}\right)r_{b}^{2}\times\nonumber \\
 & \times\int_{0}^{\infty}dr_{c}\frac{\pi\left|\mathbf{K}_{c}\right|^{2}}{2}J_{l_{c}}\left(\left|\mathbf{K}_{c}\right|r_{c}\right)R_{l_{c}}\left(r_{c}\right)r_{c}^{2}\times\int_{0}^{\infty}dr_{d}\frac{\pi\left|\mathbf{K}_{d}\right|^{2}}{2}J_{l_{d}}\left(\left|\mathbf{K}_{d}\right|r_{a}\right)R_{l_{d}}\left(r_{d}\right)r_{d}\label{eq:I_abcd-1}
\end{align}
We now perform the integrals $\int d^{3}\mathbf{r}_{1}\int d^{3}\mathbf{r}_{2}$
to obtain some delta functions that simplify the integrations:
\begin{align}
I_{abcd} & =\left(-i\right)^{l_{b}+l_{d}}i{}^{l_{a}+l_{c}}\int\frac{d^{3}\mathbf{K}}{\left(2\pi\right)^{3}}\frac{d^{3}\mathbf{K}_{a}}{\left(2\pi\right)^{3}}\frac{d^{3}\mathbf{K}_{c}}{\left(2\pi\right)^{3}}\times\nonumber \\
 & \times\frac{4\pi}{\mathbf{K}^{2}}\exp\left(i\left[\mathbf{K}_{a}\cdot\left[\mathbf{R}_{a}-\mathbf{R}_{b}\right]+\mathbf{K}_{c}\cdot\left[\mathbf{R}_{c}-\mathbf{R_{d}}\right]+\mathbf{K}\cdot\left[\mathbf{R}_{b}-\mathbf{R}_{d}\right]\right]\right)\times\nonumber \\
 & \times Y_{l_{a}m_{a}}^{*}\left(\hat{\mathbf{K}}_{a}\right)Y_{l_{c}m_{c}}^{*}\left(\hat{\mathbf{K}}_{c}\right)Y_{l_{b}m_{b}}\left(\widehat{\mathbf{K}_{a}-\mathbf{K}}\right)Y_{l_{d}m_{d}}\left(\widehat{\mathbf{K}_{c}+\mathbf{K}}\right)\times\nonumber \\
 & \times\int_{0}^{\infty}dr_{a}\frac{\pi\left|\mathbf{K}_{a}\right|^{2}}{2}J_{l_{a}}\left(\left|\mathbf{K}_{a}\right|r_{a}\right)R_{l_{a}}\left(r_{a}\right)r_{a}^{2}\times\int_{0}^{\infty}dr_{b}\frac{\pi\left|\mathbf{K}_{b}\right|^{2}}{2}J_{l_{b}}\left(\left|\mathbf{K}_{a}-\mathbf{K}\right|r_{b}\right)R_{l_{b}}\left(r_{b}\right)r_{b}^{2}\times\nonumber \\
 & \times\int_{0}^{\infty}dr_{c}\frac{\pi\left|\mathbf{K}_{c}\right|^{2}}{2}J_{l_{c}}\left(\left|\mathbf{K}_{c}\right|r_{c}\right)R_{l_{c}}\left(r_{c}\right)r_{c}^{2}\times\int_{0}^{\infty}dr_{d}\frac{\pi\left|\mathbf{K}_{d}\right|^{2}}{2}J_{l_{d}}\left(\left|\mathbf{K}_{b}+\mathbf{K}\right|r_{a}\right)R_{l_{d}}\left(r_{d}\right)r_{d}^{2}\label{eq:I_abcd-2}
\end{align}
Now we let 
\begin{equation}
F_{\alpha}\left(K\right)=\int_{0}^{\infty}dr_{\alpha}K^{2}J_{l_{\alpha}}\left(Kr_{\alpha}\right)R_{l_{\alpha}}\left(r_{\alpha}\right)r_{\alpha}^{2}\label{eq:Hard}
\end{equation}
so that 
\begin{align}
I_{abcd} & =\left(-i\right)^{l_{b}+l_{d}}i{}^{l_{a}+l_{c}}\int\frac{d^{3}\mathbf{K}}{\left(2\pi\right)^{3}}\frac{d^{3}\mathbf{K}_{a}}{\left(2\pi\right)^{3}}\frac{d^{3}\mathbf{K}_{c}}{\left(2\pi\right)^{3}}\frac{4\pi}{\mathbf{K}^{2}}\exp\left(i\left[\mathbf{K}_{a}\cdot\left[\mathbf{R}_{a}-\mathbf{R}_{b}\right]+\mathbf{K}_{c}\cdot\left[\mathbf{R}_{c}-\mathbf{R_{d}}\right]+\mathbf{K}\cdot\left[\mathbf{R}_{b}-\mathbf{R}_{d}\right]\right]\right)\times\nonumber \\
 & \times Y_{l_{a}m_{a}}^{*}\left(\hat{\mathbf{K}}_{a}\right)Y_{l_{c}m_{c}}^{*}\left(\hat{\mathbf{K}}_{c}\right)Y_{l_{b}m_{b}}\left(\widehat{\mathbf{K}_{a}-\mathbf{K}}\right)Y_{l_{d}m_{d}}\left(\widehat{\mathbf{K}_{c}+\mathbf{K}}\right)\times F_{a}\left(\left|\mathbf{K}_{a}\right|\right)F_{c}\left(\left|\mathbf{K}_{c}\right|\right)F_{b}\left(\left|\mathbf{K}_{a}-\mathbf{K}\right|\right)F_{d}\left(\left|\mathbf{K}_{b}+\mathbf{K}\right|\right)\label{eq:Good}
\end{align}

\subsection{\protect\label{subsec:Expansion-in-terms}Expansion in terms of derivatives}

Now we write
\begin{equation}
\mathcal{F}_{\alpha}\left(K\right)=\frac{F_{\alpha}\left(K\right)}{K^{l_{\alpha}}}\label{eq:rescaling}
\end{equation}
So that:
\begin{align}
I_{abcd} & =\left(-i\right)^{l_{b}+l_{d}}i{}^{l_{a}+l_{c}}\int\frac{d^{3}\mathbf{K}}{\left(2\pi\right)^{3}}\frac{d^{3}\mathbf{K}_{a}}{\left(2\pi\right)^{3}}\frac{d^{3}\mathbf{K}_{c}}{\left(2\pi\right)^{3}}\frac{4\pi}{\mathbf{K}^{2}}\exp\left(i\left[\mathbf{K}_{a}\cdot\left[\mathbf{R}_{a}-\mathbf{R}_{b}\right]+\mathbf{K}_{c}\cdot\left[\mathbf{R}_{c}-\mathbf{R_{d}}\right]+\mathbf{K}\cdot\left[\mathbf{R}_{b}-\mathbf{R}_{d}\right]\right]\right)\nonumber \\
 & \times\left|\mathbf{K}_{a}\right|^{l_{a}}Y_{l_{a}m_{a}}\left(\hat{\mathbf{K}}_{a}\right)\left|\mathbf{K}_{c}\right|^{l_{x}}Y_{l_{c}m_{c}}\left(\hat{\mathbf{K}}_{c}\right)\left|\mathbf{K}_{a}-\mathbf{K}\right|^{l_{b}}Y_{l_{b}m_{b}}^{*}\left(\widehat{\mathbf{K}_{a}-\mathbf{K}}\right)\left|\mathbf{K}_{c}+\mathbf{K}\right|^{l_{c}}Y_{l_{d}m_{d}}^{*}\left(\widehat{\mathbf{K}_{c}+\mathbf{K}}\right)\times\nonumber \\
 & \times\mathcal{F}_{a}\left(\left|\mathbf{K}_{a}\right|\right)\mathcal{F}_{c}\left(\left|\mathbf{K}_{c}\right|\right)\mathcal{F}_{b}\left(\left|\mathbf{K}_{a}-\mathbf{K}\right|\right)\mathcal{F}_{d}\left(\left|\mathbf{K}_{c}+\mathbf{K}\right|\right)\label{eq:Integral}
\end{align}
Furthermore: 
\begin{align}
I_{abcd} & =\left(-i\right)^{l_{b}+l_{d}}i{}^{l_{a}+l_{c}}\int\frac{d^{3}\mathbf{K}}{\left(2\pi\right)^{3}}\frac{d^{3}\mathbf{K}_{a}}{\left(2\pi\right)^{3}}\frac{d^{3}\mathbf{K}_{c}}{\left(2\pi\right)^{3}}\frac{4\pi}{\mathbf{K}^{2}}\exp\left(i\left[\mathbf{K}_{a}\cdot\left[\mathbf{R}_{a}-\mathbf{R}_{b}\right]+\mathbf{K}_{c}\cdot\left[\mathbf{R}_{c}-\mathbf{R_{d}}\right]+\mathbf{K}\cdot\left[\mathbf{R}_{b}-\mathbf{R}_{d}\right]\right]\right)\times\nonumber \\
 & \times\sum C_{abcd}^{m,n,p,q}\mathbf{K}_{ax}^{m_{x}}\mathbf{K}_{ay}^{m_{y}}\mathbf{K}_{az}^{m_{z}}\mathbf{K}_{cx}^{n_{x}}\mathbf{K}_{cy}^{n_{y}}\mathbf{K}_{cz}^{m_{z}}\left(\mathbf{K}_{a}-\mathbf{K}\right)_{x}^{p_{x}}\left(\mathbf{K}_{a}-\mathbf{K}\right)_{y}^{p_{y}}\left(\mathbf{K}_{a}-\mathbf{K}\right)_{z}^{p_{z}}\left(\mathbf{K}_{c}+\mathbf{K}\right)_{x}^{q_{x}}\left(\mathbf{K}_{c}+\mathbf{K}\right)_{y}^{q_{y}}\left(\mathbf{K}_{c}+\mathbf{K}\right)_{z}^{q_{z}}\times\nonumber \\
 & \times\mathcal{F}_{a}\left(\left|\mathbf{K}_{a}\right|\right)\mathcal{F}_{c}\left(\left|\mathbf{K}_{c}\right|\right)\mathcal{F}_{b}\left(\left|\mathbf{K}_{a}-\mathbf{K}\right|\right)\mathcal{F}_{d}\left(\left|\mathbf{K}_{c}+\mathbf{K}\right|\right)\label{eq:Integral_II}
\end{align}
Where we have transformed spherical harmonic into solid harmonics.
Then 
\begin{align}
I_{abcd} & =\left(-i\right)^{l_{b}+l_{d}}i{}^{l_{a}+l_{c}}\sum D_{abcd}^{m,n,p}\frac{\partial^{m_{x}}}{\partial\mathbf{R}_{a}^{x}}\frac{\partial^{m_{y}}}{\partial\mathbf{R}_{a}^{y}}\frac{\partial^{m_{z}}}{\partial\mathbf{R}_{a}^{z}}\frac{\partial^{n_{x}}}{\partial\mathbf{R}_{c}^{x}}\frac{\partial^{n_{y}}}{\partial\mathbf{R}_{c}^{y}}\frac{\partial^{n_{z}}}{\partial\mathbf{R}_{c}^{z}}\frac{\partial^{p_{x}}}{\partial\mathbf{R}_{b}^{x}}\frac{\partial^{p_{y}}}{\partial\mathbf{R}_{b}^{y}}\frac{\partial^{p_{z}}}{\partial\mathbf{R}_{b}^{z}}\times\nonumber \\
 & \times\int\frac{d^{3}\mathbf{K}}{\left(2\pi\right)^{3}}\frac{d^{3}\mathbf{K}_{a}}{\left(2\pi\right)^{3}}\frac{d^{3}\mathbf{K}_{c}}{\left(2\pi\right)^{3}}\frac{4\pi}{\mathbf{K}^{2}}\exp\left(i\left[\mathbf{K}_{a}\cdot\left[\mathbf{R}_{a}-\mathbf{R}_{b}\right]+\mathbf{K}_{c}\cdot\left[\mathbf{R}_{c}-\mathbf{R}_{d}\right]+\mathbf{K}\cdot\left[\mathbf{R}_{b}-\mathbf{R}_{d}\right]\right]\right)\times\nonumber \\
 & \times\mathcal{F}_{a}\left(\left|\mathbf{K}_{a}\right|\right)\mathcal{F}_{c}\left(\left|\mathbf{K}_{c}\right|\right)\mathcal{F}_{b}\left(\left|\mathbf{K}_{a}-\mathbf{K}\right|\right)\mathcal{F}_{d}\left(\left|\mathbf{K}_{c}+\mathbf{K}\right|\right)\label{eq:Intgeral_III}
\end{align}
Where we have used the derivative property of exponentials to pull
out the solid harmonics as derivatives. Whereby: 
\begin{align}
I_{abcd} & =\left(-i\right)^{l_{b}+l_{d}}i{}^{l_{a}+l_{c}}\sum D_{abcd}^{m,n,p}\frac{\partial^{m_{x}}}{\partial\mathbf{R}_{a}^{x}}\frac{\partial^{m_{y}}}{\partial\mathbf{R}_{a}^{y}}\frac{\partial^{m_{z}}}{\partial\mathbf{R}_{a}^{z}}\frac{\partial^{n_{x}}}{\partial\mathbf{R}_{c}^{x}}\frac{\partial^{n_{y}}}{\partial\mathbf{R}_{c}^{y}}\frac{\partial^{n_{z}}}{\partial\mathbf{R}_{c}^{z}}\frac{\partial^{p_{x}}}{\partial\mathbf{R}_{b}^{x}}\frac{\partial^{p_{y}}}{\partial\mathbf{R}_{b}^{y}}\frac{\partial^{p_{z}}}{\partial\mathbf{R}_{b}^{z}}\times\nonumber \\
 & \times\int\frac{d^{3}\mathbf{K}}{\left(2\pi\right)^{3}}\frac{d^{3}\mathbf{K}_{a}}{\left(2\pi\right)^{3}}\frac{d^{3}\mathbf{K}_{c}}{\left(2\pi\right)^{3}}\frac{4\pi}{\mathbf{K}^{2}}\exp\left(i\left[\mathbf{K}_{a}\cdot\left[\mathbf{R}_{a}-\mathbf{R}_{b}\right]+\mathbf{K}_{c}\cdot\left[\mathbf{R}_{c}-\mathbf{R}_{d}\right]+\mathbf{K}\cdot\left[\mathbf{R}_{b}-\mathbf{R}_{d}\right]\right]\right)\times\nonumber \\
 & \times\mathcal{F}_{a}\left(\left|\mathbf{K}_{a}\right|\right)\mathcal{F}_{c}\left(\left|\mathbf{K}_{c}\right|\right)\mathcal{F}_{b}\left(\left|\mathbf{K}_{a}-\mathbf{K}\right|\right)\mathcal{F}_{d}\left(\left|\mathbf{K}_{c}+\mathbf{K}\right|\right)\label{eq:Integral_IV}
\end{align}

\subsection{\protect\label{subsec:A-Bessel-function}A Bessel function expansion}
Now \citep{Louks_1967}:

\begin{equation}
\exp\left(i\mathbf{K}_{a}\left(\mathbf{R}_{a}-\mathbf{R}_{b}\right)\right)=\sum_{L_{a}M_{a}}i^{L_{a}}J_{L_{a}}\left(\left|\mathbf{K}_{a}\right|\left|\mathbf{R}_{a}-\mathbf{R}_{b}\right|\right)Y_{L_{a}M_{a}}^{*}\left(\hat{\mathbf{K}}_{\alpha}\right)Y_{L_{a}M_{a}}\left(\widehat{\mathbf{R}_{a}-\mathbf{\mathbf{R}}_{b}}\right)\label{eq:Substitution}
\end{equation}
As such: 
\begin{align}
I_{abcd} & =\left(-i\right)^{l_{b}+l_{d}}i{}^{l_{a}+l_{c}}\sum D_{abcd}^{m,n,p}\frac{\partial^{m_{x}}}{\partial\mathbf{R}_{a}^{x}}\frac{\partial^{m_{y}}}{\partial\mathbf{R}_{a}^{y}}\frac{\partial^{m_{z}}}{\partial\mathbf{R}_{a}^{z}}\frac{\partial^{n_{x}}}{\partial\mathbf{R}_{c}^{x}}\frac{\partial^{n_{y}}}{\partial\mathbf{R}_{c}^{y}}\frac{\partial^{n_{z}}}{\partial\mathbf{R}_{c}^{z}}\frac{\partial^{p_{x}}}{\partial\mathbf{R}_{b}^{x}}\frac{\partial^{p_{y}}}{\partial\mathbf{R}_{b}^{y}}\frac{\partial^{p_{z}}}{\partial\mathbf{R}_{b}^{z}}\times\nonumber \\
 & \sum_{L_{a}M_{a}}\sum_{L_{c},M_{c}}\sum_{L,M}i^{L_{a}}Y_{L_{a}M_{a}}\left(\widehat{\mathbf{R}_{a}-\mathbf{\mathbf{R}}_{b}}\right)i^{L_{c}}Y_{L_{c}M_{c}}\left(\widehat{\mathbf{R}_{c}-\mathbf{\mathbf{R}}_{d}}\right)i^{L}Y_{LM}\left(\widehat{\mathbf{R}_{b}-\mathbf{\mathbf{R}}_{d}}\right)\nonumber \\
 & \times\int\frac{d^{3}\mathbf{K}}{\left(2\pi\right)^{3}}\frac{d^{3}\mathbf{K}_{a}}{\left(2\pi\right)^{3}}\frac{d^{3}\mathbf{K}_{c}}{\left(2\pi\right)^{3}}\frac{4\pi}{\mathbf{K}^{2}}J_{L_{a}}\left(\left|\mathbf{K}_{a}\right|\left|\mathbf{R}_{a}-\mathbf{R}_{b}\right|\right)J_{L_{c}}\left(\left|\mathbf{K}_{c}\right|\left|\mathbf{R}_{c}-\mathbf{R}_{d}\right|\right)J_{L}\left(\left|\mathbf{K}\right|\left|\mathbf{R}_{b}-\mathbf{R}_{d}\right|\right)\nonumber \\
 & \times\mathcal{F}_{a}\left(\left|\mathbf{K}_{a}\right|\right)\mathcal{F}_{c}\left(\left|\mathbf{K}_{c}\right|\right)\mathcal{F}_{b}\left(\left|\mathbf{K}_{a}-\mathbf{K}\right|\right)\mathcal{F}_{d}\left(\left|\mathbf{K}_{c}+\mathbf{K}\right|\right)\label{eq:Integral_V}
\end{align}
and 
\begin{align}
I_{abcd} & =4\pi\left(-i\right)^{l_{b}+l_{d}}i{}^{l_{a}+l_{c}}\sum D_{abcd}^{m,n,p}\frac{\partial^{m_{x}}}{\partial\mathbf{R}_{a}^{x}}\frac{\partial^{m_{y}}}{\partial\mathbf{R}_{a}^{y}}\frac{\partial^{m_{z}}}{\partial\mathbf{R}_{a}^{z}}\frac{\partial^{n_{x}}}{\partial\mathbf{R}_{c}^{x}}\frac{\partial^{n_{y}}}{\partial\mathbf{R}_{c}^{y}}\frac{\partial^{n_{z}}}{\partial\mathbf{R}_{c}^{z}}\frac{\partial^{p_{x}}}{\partial\mathbf{R}_{b}^{x}}\frac{\partial^{p_{y}}}{\partial\mathbf{R}_{b}^{y}}\frac{\partial^{p_{z}}}{\partial\mathbf{R}_{b}^{z}}\times\nonumber \\
 & \times\sum_{L_{a}M_{a}}\sum_{L_{c},M_{c}}\sum_{L,M}i^{L_{a}}Y_{L_{a}M_{a}}\left(\widehat{\mathbf{R}_{a}-\mathbf{\mathbf{R}}_{b}}\right)i^{L_{c}}Y_{L_{c}M_{c}}\left(\widehat{\mathbf{R}_{c}-\mathbf{\mathbf{R}}_{d}}\right)i^{L}Y_{LM}\left(\widehat{\mathbf{R}_{b}-\mathbf{\mathbf{R}}_{d}}\right)\nonumber \\
 & \times\int\frac{d\left|\mathbf{K}\right|}{2\pi}\frac{d\left|\mathbf{K}_{a}\right|du_{a}}{\left(2\pi\right)^{2}}\frac{d\left|\mathbf{K}_{c}\right|du_{c}}{\left(2\pi\right)^{2}}\times J_{L_{a}}\left(\left|\mathbf{K}_{a}\right|\left|\mathbf{R}_{a}-\mathbf{R}_{b}\right|\right)J_{L_{c}}\left(\left|\mathbf{K}_{c}\right|\left|\mathbf{R}_{c}-\mathbf{R}_{d}\right|\right)J_{L}\left(\left|\mathbf{K}\right|\left|\mathbf{R}_{b}-\mathbf{R}_{d}\right|\right)\nonumber \\
 & \times\mathcal{G}_{a}\left(\mathbf{K}_{a}^{2}\right)\mathcal{G}_{c}\left(\mathbf{K}_{c}^{2}\right)\mathcal{G}_{b}\left(\mathbf{K}_{a}^{2}+\mathbf{K}^{2}-2\left|\mathbf{K}_{a}\right|\left|\mathbf{K}\right|u_{a}\right)\mathcal{G}_{d}\left(\mathbf{K}_{c}^{2}+\mathbf{K}^{2}-2\left|\mathbf{K}_{c}\right|\left|\mathbf{K}\right|u_{c}\right)\left|\mathbf{K}_{a}\right|^{2}\left|\mathbf{K}_{c}\right|^{2}\label{eq:Integral_VI}
\end{align}
The extra minus in $\mathcal{G}_{d}\left(\mathbf{K}_{c}^{2}+\mathbf{K}^{2}-\left|\mathbf{K}_{c}\right|\left|\mathbf{K}\right|u_{c}\right)$
is because we are integrating $u_{c}$ between $\left[-1,1\right]$.
Where 
\begin{equation}
\mathcal{G}\left(K^{2}\right)=\mathcal{F}\left(K\right)=\frac{F_{\alpha}\left(K\right)}{K^{l_{\alpha}}}=\frac{\int_{0}^{\infty}dr_{\alpha}J_{l_{\alpha}}\left(Kr_{\alpha}\right)R_{l_{\alpha}}\left(r_{\alpha}\right)r_{\alpha}^{2}}{K^{l_{\alpha}}}\label{eq:easier}
\end{equation}
Now we write: 
\begin{align}
I_{abcd} & =4\pi\left(-i\right)^{l_{b}+l_{d}}i{}^{l_{a}+l_{c}}\sum D_{abcd}^{m,n,p}\frac{\partial^{m_{x}}}{\partial\mathbf{R}_{a}^{x}}\frac{\partial^{m_{y}}}{\partial\mathbf{R}_{a}^{y}}\frac{\partial^{m_{z}}}{\partial\mathbf{R}_{a}^{z}}\frac{\partial^{n_{x}}}{\partial\mathbf{R}_{c}^{x}}\frac{\partial^{n_{y}}}{\partial\mathbf{R}_{c}^{y}}\frac{\partial^{n_{z}}}{\partial\mathbf{R}_{c}^{z}}\frac{\partial^{p_{x}}}{\partial\mathbf{R}_{b}^{x}}\frac{\partial^{p_{y}}}{\partial\mathbf{R}_{b}^{y}}\frac{\partial^{p_{z}}}{\partial\mathbf{R}_{b}^{z}}\times\nonumber \\
 & \times\sum_{L_{a}M_{a}}\sum_{L_{c},M_{c}}\sum_{L,M}i^{L_{a}}Y_{L_{a}M_{a}}\left(\widehat{\mathbf{R}_{a}-\mathbf{\mathbf{R}}_{b}}\right)i^{L_{c}}Y_{L_{c}M_{c}}\left(\widehat{\mathbf{R}_{c}-\mathbf{\mathbf{R}}_{d}}\right)i^{L}Y_{LM}\left(\widehat{\mathbf{R}_{b}-\mathbf{\mathbf{R}}_{d}}\right)\times\nonumber \\
 & \times\int\frac{d\left|\mathbf{K}\right|}{2\pi}\frac{d\left|\mathbf{K}_{a}\right|du_{a}}{\left(2\pi\right)^{2}}\frac{d\left|\mathbf{K}_{c}\right|du_{c}}{\left(2\pi\right)^{2}}\times J_{L_{a}}\left(\left|\mathbf{K}_{a}\right|\left|\mathbf{R}_{a}-\mathbf{R}_{b}\right|\right)J_{L_{c}}\left(\left|\mathbf{K}_{c}\right|\left|\mathbf{R}_{c}-\mathbf{R}_{d}\right|\right)J_{L}\left(\left|\mathbf{K}\right|\left|\mathbf{R}_{b}-\mathbf{R}_{d}\right|\right)\times\nonumber \\
 & \times\frac{\int_{0}^{\infty}dr_{a}J_{l_{a}}\left(\left|\mathbf{K}_{a}\right|r_{a}\right)R_{l_{a}}\left(r_{a}\right)r_{a}^{2}}{\left|\mathbf{K}_{a}\right|^{l_{a}-2}}\frac{\int_{0}^{\infty}dr_{c}J_{l_{c}}\left(\left|\mathbf{K}_{c}\right|r_{c}\right)R_{l_{c}}\left(r_{c}\right)r_{c}^{2}}{\left|\mathbf{K}_{c}\right|^{l_{c}-2}}\times\nonumber \\
 & \times\frac{\int_{0}^{\infty}dr_{b}J_{l_{b}}\left(\left[\mathbf{K}_{a}^{2}+\mathbf{K}^{2}-2\left|\mathbf{K}_{a}\right|\left|\mathbf{K}\right|u_{a}\right]^{1/2}r_{b}\right)R_{l_{b}}\left(r_{b}\right)r_{b}^{2}}{\left[\mathbf{K}_{a}^{2}+\mathbf{K}^{2}-2\left|\mathbf{K}_{a}\right|\left|\mathbf{K}\right|u_{a}\right]^{l_{b}/2}}\frac{\int_{0}^{\infty}dr_{d}J_{l_{d}}\left(\left[\mathbf{K}_{c}^{2}+\mathbf{K}^{2}-2\left|\mathbf{K}_{c}\right|\left|\mathbf{K}\right|u_{c}\right]^{1/2}r_{d}\right)R_{l_{d}}\left(r_{d}\right)r_{d}^{2}}{\left[\mathbf{K}_{c}^{2}+\mathbf{K}^{2}-2\left|\mathbf{K}_{b}\right|\left|\mathbf{K}\right|u_{c}\right]^{l_{d}/2}}\label{eq:Expansion}
\end{align}
\end{widetext}

\subsection{\protect\label{subsec:A-global-co-ordinate}A global co-ordinate
change}

Now we use the co-ordinates: 
\begin{align}
K_{a} & =\mathbf{K}_{a}\nonumber \\
K_{c} & =\mathbf{K}_{c}\nonumber \\
K & =\mathbf{K}\nonumber \\
U_{a} & =\left[\mathbf{K}_{a}^{2}+\mathbf{K}^{2}-2\left|\mathbf{K}_{a}\right|\left|\mathbf{K}\right|u_{a}\right]^{1/2}\nonumber \\
U_{c} & =\left[\mathbf{K}_{c}^{2}+\mathbf{K}^{2}-2\left|\mathbf{K}_{c}\right|\left|\mathbf{K}\right|u_{c}\right]^{1/2}\label{eq:co-ordinates}
\end{align}
Now we have that the matrix of the transformation is upper triangular
so the determinant of the Jacobian is given by:
\begin{equation}
\det\left(\mathcal{J}\right)=\frac{K^{2}K_{a}K_{c}}{U_{a}U_{c}}\label{eq:Jacobian}
\end{equation}
As such we have that: 
\begin{widetext}
\begin{align}
I_{abcd} & =4\pi\left(-i\right)^{l_{b}+l_{d}}i{}^{l_{a}+l_{c}}\sum D_{abcd}^{m,n,p}\frac{\partial^{m_{x}}}{\partial\mathbf{R}_{a}^{x}}\frac{\partial^{m_{y}}}{\partial\mathbf{R}_{a}^{y}}\frac{\partial^{m_{z}}}{\partial\mathbf{R}_{a}^{z}}\frac{\partial^{n_{x}}}{\partial\mathbf{R}_{c}^{x}}\frac{\partial^{n_{y}}}{\partial\mathbf{R}_{c}^{y}}\frac{\partial^{n_{z}}}{\partial\mathbf{R}_{c}^{z}}\frac{\partial^{p_{x}}}{\partial\mathbf{R}_{b}^{x}}\frac{\partial^{p_{y}}}{\partial\mathbf{R}_{b}^{y}}\frac{\partial^{p_{z}}}{\partial\mathbf{R}_{b}^{z}}\times\nonumber \\
 & \times\sum_{L_{a}M_{a}}\sum_{L_{c},M_{c}}\sum_{L,M}i^{L_{a}}Y_{L_{a}M_{a}}\left(\widehat{\mathbf{R}_{a}-\mathbf{\mathbf{R}}_{b}}\right)i^{L_{c}}Y_{L_{c}M_{c}}\left(\widehat{\mathbf{R}_{c}-\mathbf{\mathbf{R}}_{d}}\right)i^{L}Y_{LM}\left(\widehat{\mathbf{R}_{b}-\mathbf{\mathbf{R}}_{d}}\right)\times\nonumber \\
 & \times\int\frac{1}{\left(2\pi\right)^{5}}\frac{U_{a}U_{c}}{K^{2}K_{a}K_{c}}\times J_{L_{a}}\left(K_{a}\left|\mathbf{R}_{a}-\mathbf{R}_{b}\right|\right)J_{L_{c}}\left(K_{c}\left|\mathbf{R}_{c}-\mathbf{R}_{d}\right|\right)J_{L}\left(K\left|\mathbf{R}_{b}-\mathbf{R}_{d}\right|\right)\times\nonumber \\
 & \times\left(\frac{2}{\pi}\right)^{4}\frac{\int_{0}^{\infty}dr_{a}J_{l_{a}}\left(K_{a}r_{a}\right)R_{l_{a}}\left(r_{a}\right)r_{a}^{2}}{K_{a}^{l_{a}}}\frac{\int_{0}^{\infty}dr_{c}J_{l_{c}}\left(K_{c}r_{c}\right)R_{l_{c}}\left(r_{c}\right)r_{c}^{2}}{K_{c}^{l_{c}}}\times\nonumber \\
 & \times\frac{\int_{0}^{\infty}dr_{b}J_{l_{b}}\left(U_{a}r_{b}\right)R_{l_{b}}\left(r_{b}\right)r_{b}^{2}}{U_{a}^{l_{b}+2}}\frac{\int_{0}^{\infty}dr_{d}J_{l_{d}}\left(U_{c}r_{d}\right)R_{l_{d}}\left(r_{d}\right)r_{d}^{2}}{U_{c}^{l_{d}+2}}\label{eq:Expansion-1}
\end{align}
This factorizes to:
\begin{align}
I_{abcd} & =4\pi\left(-i\right)^{l_{b}+l_{d}}i{}^{l_{a}+l_{c}}\sum D_{abcd}^{m,n,p}\frac{\partial^{m_{x}}}{\partial\mathbf{R}_{a}^{x}}\frac{\partial^{m_{y}}}{\partial\mathbf{R}_{a}^{y}}\frac{\partial^{m_{z}}}{\partial\mathbf{R}_{a}^{z}}\frac{\partial^{n_{x}}}{\partial\mathbf{R}_{c}^{x}}\frac{\partial^{n_{y}}}{\partial\mathbf{R}_{c}^{y}}\frac{\partial^{n_{z}}}{\partial\mathbf{R}_{c}^{z}}\frac{\partial^{p_{x}}}{\partial\mathbf{R}_{b}^{x}}\frac{\partial^{p_{y}}}{\partial\mathbf{R}_{b}^{y}}\frac{\partial^{p_{z}}}{\partial\mathbf{R}_{b}^{z}}\times\nonumber \\
 & \times\sum_{L_{a}M_{a}}\sum_{L_{c},M_{c}}\sum_{L,M}i^{L_{a}}Y_{L_{a}M_{a}}\left(\widehat{\mathbf{R}_{a}-\mathbf{\mathbf{R}}_{b}}\right)i^{L_{c}}Y_{L_{c}M_{c}}\left(\widehat{\mathbf{R}_{c}-\mathbf{\mathbf{R}}_{d}}\right)i^{L}Y_{LM}\left(\widehat{\mathbf{R}_{b}-\mathbf{\mathbf{R}}_{d}}\right)\nonumber \\
 & \times\int\frac{dK_{a}}{4}\frac{\int_{0}^{\infty}dr_{a}J_{l_{a}}\left(K_{a}r_{a}\right)R_{l_{a}}\left(r_{a}\right)r_{a}^{2}}{K_{a}^{l_{a}+1}}J_{L_{a}}\left(K_{a}\left|\mathbf{R}_{a}-\mathbf{R}_{b}\right|\right)\nonumber \\
 & \times\int_{0}^{\infty}\frac{dK_{c}}{4}\frac{\int_{0}^{\infty}dr_{c}J_{l_{c}}\left(K_{c}r_{c}\right)R_{l_{c}}\left(r_{c}\right)r_{c}^{2}}{K_{c}^{l_{c}+1}}J_{L_{c}}\left(K_{c}\left|\mathbf{R}_{c}-\mathbf{R}_{d}\right|\right)\nonumber \\
 & \times\int_{0}^{\infty}\frac{dK}{2\pi}\frac{1}{K^{2}}J_{L}\left(K\left|\mathbf{R}_{b}-\mathbf{R}_{d}\right|\right)\nonumber \\
 & \times\int_{\left|K_{a}-K\right|}^{K_{a}+K}\frac{dU_{a}}{4}\frac{\int_{0}^{\infty}dr_{b}J_{l_{b}}\left(U_{a}r_{b}\right)R_{l_{b}}\left(r_{b}\right)r_{b}^{2}}{U_{a}^{l_{b}+1}}\nonumber \\
 & \times\int_{\left|K_{c}-K\right|}^{K_{c}+K}\frac{dU_{c}}{4}\frac{\int_{0}^{\infty}dr_{d}J_{l_{b}}\left(U_{c}r_{d}\right)R_{l_{d}}\left(r_{d}\right)r_{d}^{2}}{U_{c}^{l_{d}+1}}\label{eq:I_abcd}
\end{align}

\end{widetext}
This is our main result as it presents the Coulomb four center integrals
in a nearly factorable form. 

\section{\protect\label{sec:Conclusions}Conclusions}

In this work we have presented some formulas for four center integrals
associated with the Coulomb interaction. These results are in a nearly
factorable form and depend on the Fourier Bessel transforms of the
radial wave function given in Eq. (\ref{eq:Bessel_transform-1}),
as well as auxiliary integrals see Eq. (\ref{eq:I_abcd}). In the
future it would be of interest to combine the new basis sets introduced
in Ref. \citep{Goldstein_2024(2)} where all basis wave functions
are of the form considered in Eq. (\ref{eq:Wave_function}) to study
the electronic structures of small molecules through either HF or
DFT methods.

\end{document}